\documentclass[conference]{IEEEtran}
\IEEEoverridecommandlockouts

\usepackage{amsmath,amssymb,amsfonts}
\usepackage{algorithm}
\usepackage{algorithmic}
\usepackage{graphicx}
\usepackage{textcomp}
\usepackage{xcolor}
\usepackage{booktabs}
\usepackage{multirow}
\usepackage{url}
\usepackage{placeins}
\usepackage{listings}
\usepackage[colorlinks=true,linkcolor=black,citecolor=black,urlcolor=black,bookmarks=true]{hyperref}
\usepackage{cite}

\makeatletter
\renewcommand{\footnoterule}{\kern-3\p@\hrule\@width 0.4\columnwidth\kern 2.6\p@}
\makeatother

\def\BibTeX{{\rm B\kern-.05em{\sc i\kern-.025em b}\kern-.08em
    T\kern-.1667em\lower.7ex\hbox{E}\kern-.125emX}}

\begin{document}

\title{KnowFeat: Knowledge-Guided Feature Engineering via LLM Agents}

\author{
\IEEEauthorblockN{Chengsong You$^{1}$,
Wangyue Li$^{1}$,
Weiqiao Que$^{1}$,
Qizhou Chen$^{1}$,\\
Kunyan Wu$^{2}$,
Wei Deng$^{3}$,
Feng Zhu$^{4}$,
Xiaofeng He$^{1,*}$\thanks{$^{*}$Corresponding author.}}
\IEEEauthorblockA{$^{1}$East China Normal University, Shanghai, China\\
\{51275901122, 52285901017\}@stu.ecnu.edu.cn, weiqiaoque@gmail.com,\\
582476065@qq.com, hexf@cs.ecnu.edu.cn}
\IEEEauthorblockA{$^{2}$Tsinghua University, Beijing, China\\
wky24@mails.tsinghua.edu.cn}
\IEEEauthorblockA{$^{3}$Southwestern University of Finance and Economics, Chengdu, China\\
dengwei@swufe.edu.cn}
\IEEEauthorblockA{$^{4}$Digital Currency Department, Postal Savings Bank of China, Beijing, China\\
\href{mailto:Jason_bank@126.com}{Jason\_bank@126.com}}
}

\maketitle

% Abstract
% abstract.tex
\begin{abstract}
Automated feature engineering with large language models (LLMs) can produce semantically meaningful features for tabular data, yet existing methods lack structured domain knowledge, rigorous verification, and explainable provenance. We propose KnowFeat, a knowledge-guided feature engineering framework that organizes domain knowledge into five types---schema metadata, regulatory indicators, detection rules, expert opinions, and court document evidence---and injects them as structured context into an LLM agent. A three-stage verification pipeline filters candidates through code execution, statistical quality checks, and model effectiveness evaluation. Every accepted feature carries a provenance record tracing its design to specific knowledge assets. Under a strict held-out protocol that eliminates feature-selection leakage, KnowFeat ranks first (avg.\ rank 2.3) across twelve public benchmarks among seven methods (one-sided Wilcoxon $p{=}0.017$), with a peak gain of +11.6\,pp AUC on a telecom churn dataset. On a real-world Bitcoin anti-money laundering (AML) dataset (Elliptic) and a synthetic digital currency AML benchmark (SimECNY), KnowFeat maintains competitive detection performance with full provenance traceability.
\end{abstract}

% Introduction
% intro.tex
\section{Introduction}
\label{sec:intro}

Feature engineering is among the most impactful steps in building machine learning models for tabular data~\cite{domingos2012few}. Good features improve accuracy and make models easier to interpret---but designing them by hand requires domain expertise that is scarce and does not transfer between applications.

Traditional automated feature engineering (AutoFE) methods such as OpenFE~\cite{zhang2023openfe} and AutoFeat~\cite{horn2019autofeat} apply systematic transformations (mathematical operations, aggregations, and feature crosses) to existing columns, then select the most predictive ones. These methods are purely data-driven: they explore a predefined transformation space without understanding what the features \emph{mean}. In regulated domains such as finance and healthcare, stakeholders must know \emph{what} a feature represents before deploying it---semantic meaning is not optional.

The emergence of large language models (LLMs) has opened a new paradigm for feature engineering. Methods such as CAAFE~\cite{hollmann2023caafe}, FeatLLM~\cite{han2024featllm}, and LLM-FE~\cite{abhyankar2025llmfe} use LLMs to generate feature transformation code from dataset descriptions, leveraging the model's broad world knowledge to propose semantically meaningful features. These methods show gains on general benchmarks but were not designed for regulated, domain-specific settings. Three limitations stand out:

\textbf{(1) Lack of structured domain knowledge.}
LLM-based methods rely on the model's parametric knowledge, which is often outdated or absent for specialized domains such as anti-money laundering (AML) or clinical risk scoring. Rich external knowledge sources exist---regulatory indicators, detection rules, expert opinions, court evidence---but no existing method organizes them into a structured taxonomy for systematic injection into feature generation.

\textbf{(2) Insufficient feature verification.}
Even when useful candidate features are generated, verification is weak. CAAFE relies on a logistic regression proxy; LLM-FE uses downstream accuracy without statistical filtering; FeatLLM performs no model-based validation at all. Unverified features---syntactically valid but statistically redundant or model-harmful---enter the training data unchecked.

\textbf{(3) Limited explainability.}
In regulated industries, model features must be both explainable and auditable. Decision-makers need to know \emph{why} a feature was created and \emph{what evidence} supports it. No existing method provides structured provenance linking each feature to specific knowledge assets.

We propose \textbf{KnowFeat} (\textbf{Know}ledge-guided \textbf{Feat}ure engineering), a framework that tackles all three:

\begin{itemize}
    \item \textbf{Multi-Type Knowledge Injection.} Domain knowledge is organized into five types of increasing specificity (schema metadata, risk indicators, detection rules, expert opinions, and court document evidence) and injected as structured context into the LLM agent.

    \item \textbf{Three-Stage Verification Pipeline.} Each candidate passes through L1 (code execution sandbox), L2 (statistical quality with noise-calibrated thresholds), and L3 (model effectiveness via XGBoost across 5 seeds). Early stages are cheap; expensive evaluation runs only on candidates that survive.

    \item \textbf{Explainable Provenance.} Each accepted feature carries a provenance card recording the knowledge assets, reasoning chain, computation steps, and verification outputs.

\end{itemize}

The source code\footnote{\url{https://github.com/youchengsong/large-llm-create-feature}} and datasets\footnote{\url{https://huggingface.co/datasets/anonymous-7219/knowfeat}} are publicly accessible.

We evaluate KnowFeat on twelve public tabular datasets, a synthetic AML benchmark (SimECNY, 200K transactions), and a real-world Bitcoin AML dataset (Elliptic, 46K labeled transactions). All experiments use a strict held-out protocol that eliminates the feature-selection leakage present in prior work. Key findings:

\begin{itemize}
    \item On public datasets, KnowFeat ranks first (avg.\ rank 2.3) among seven methods under leakage-free evaluation, with a significant one-sided Wilcoxon test ($p{=}0.017$). The largest gain is +11.6\,pp AUC on a telecom churn dataset.
    \item On both AML datasets, KnowFeat's verified features remain competitive with strong baselines; on SimECNY, FeatLLM's 1{,}124-feature set lowers F1, showing that volume without verification hurts.
    \item Every accepted feature carries a provenance record linking it to specific knowledge assets, enabling audit in regulated domains.
\end{itemize}

% Related Work
% related_work.tex
\section{Related Work}
\label{sec:related}

% Framework figure - placed early so it appears at top of page 3, before Table I
\begin{figure*}[!t]
    \centering
    \includegraphics[width=\textwidth]{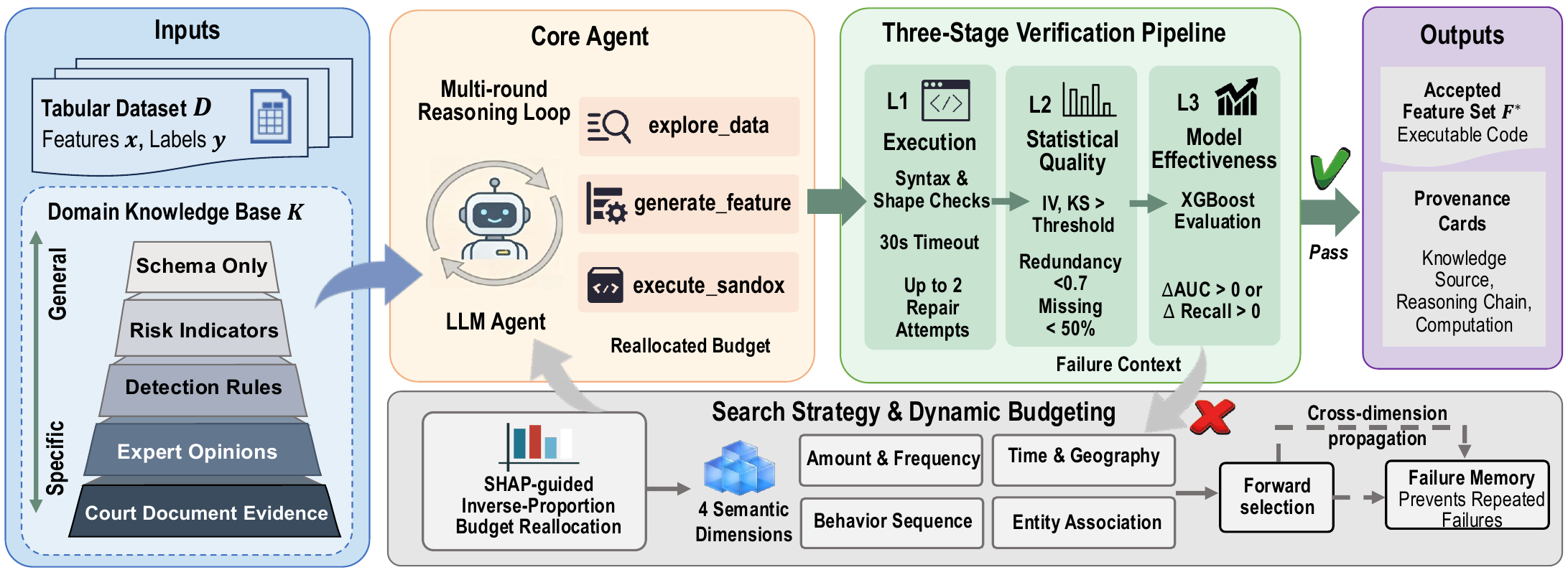}
    \caption{Overview of the KnowFeat framework. The LLM agent generates features guided by structured domain knowledge. Each candidate is verified through a three-stage pipeline (L1--L3). SHAP feedback dynamically reallocates the exploration budget across semantic dimensions.}
    \label{fig:framework}
\end{figure*}

\subsection{Traditional Automated Feature Engineering}

Traditional AutoFE methods search a \emph{syntactic} transformation space. Deep Feature Synthesis~\cite{kanter2015deep} applies aggregation primitives along relational paths; ExploreKit~\cite{katz2016explorekit} enumerates operators and ranks candidates via meta-learning; AutoFeat~\cite{horn2019autofeat} uses polynomial expansion with L1 selection; and OpenFE~\cite{zhang2023openfe} frames feature generation as an operator-operand search with LightGBM-based importance ranking. End-to-end AutoML systems such as Auto-sklearn~\cite{feurer2015auto} and FLAML~\cite{wang2021flaml} include feature preprocessing but do not generate semantically new features. These methods apply predefined mathematical operators without \emph{semantic} understanding: \texttt{age $\times$ income} is treated identically to \texttt{age $\times$ zip\_code}, even though only the former has financial meaning. They cannot generate domain-specific features (e.g., ``transaction velocity'' in anomaly detection) that require conceptual understanding~\cite{nargesian2017learning}.

\subsection{LLM-Based Feature Engineering}

Large language models~\cite{brown2020gpt3,touvron2023llama} have enabled a shift from syntactic to semantic feature generation. CAAFE~\cite{hollmann2023caafe} iteratively prompts an LLM with dataset descriptions and evaluation feedback to generate Python code for new features, using logistic regression as a proxy evaluator. FeatLLM~\cite{han2024featllm} takes a two-pass approach: first extracting natural language rules from the LLM, then converting these rules into binary indicator features. LLM-FE~\cite{abhyankar2025llmfe} frames feature engineering as an evolutionary optimization problem, using LLMs to iteratively generate and refine features through structured prompts with model performance feedback. OCTree~\cite{nam2024octree} guides LLM-based feature generation with decision tree reasoning, using tree structure feedback to steer the generation process. In the financial domain, RiskTagger~\cite{li2025risktagger} uses an LLM agent with regulatory knowledge to annotate Web3 crypto laundering behaviors, demonstrating that domain knowledge can enhance LLM-based data annotation.

These methods vary in their knowledge use and verification. LLM-FE incorporates limited domain context (dataset descriptions), but none organizes knowledge into a structured taxonomy for systematic injection. Verification also varies: CAAFE uses a weak proxy model, LLM-FE uses downstream model accuracy, OCTree uses decision tree feedback, and FeatLLM skips model-based validation entirely. None combines code execution, statistical filtering, and model evaluation in a single pipeline, nor traces each feature back to specific knowledge assets.

Tool-augmented LLM agents~\cite{yao2023react} have been applied to code generation~\cite{qian2024chatdev}, software engineering~\cite{jimenez2024swe}, data analysis~\cite{hong2024data}, and scientific discovery~\cite{boiko2023autonomous}. In data science, AIDE~\cite{jiang2025aide} and DS-Agent~\cite{guo2024ds} automate end-to-end pipelines but do not address knowledge-grounded feature generation. KnowFeat builds on the tool-use paradigm by equipping an LLM agent with specialized tools for data exploration, feature generation with provenance, and multi-stage verification.

Table~\ref{tab:comparison} summarizes the differences. KnowFeat is, to our knowledge, the first LLM-based feature engineering method that combines structured domain knowledge injection, multi-stage verification, and feature provenance tracking.

\subsection{Domain Knowledge Integration in Machine Learning}

Knowledge-infused learning~\cite{kursuncu2020knowledge} embeds external knowledge graphs into neural architectures, and transfer learning~\cite{weiss2016survey} transfers knowledge through pretrained representations, but both target unstructured data rather than tabular features. In financial AML, domain experts manually encode risk indicators, transaction structuring patterns, and network centrality measures as features~\cite{chen2018aml}. Regulatory bodies publish red-flag typologies (e.g., FATF guidelines~\cite{fatf2012recommendations}), and institutions maintain internal detection rules from compliance experience and legal proceedings~\cite{weber2019anti}. This rich, structured knowledge has not been tapped by automated feature engineering methods.

\begin{table}[t]
\centering
\caption{Comparison of automated feature engineering methods. \checkmark: fully supported; $\circ$: partially supported; blank: not supported; --: not applicable.}
\label{tab:comparison}
\small
\renewcommand{\arraystretch}{1.2}
\setlength{\tabcolsep}{1pt}
\begin{tabular}{l c c c c c c c}
\hline
                   & \textbf{Open} & \textbf{Auto} & \textbf{CAA} & \textbf{Feat} & \textbf{LLM-} & \textbf{OC} & \textbf{Know} \\
\textbf{Property}  & \textbf{FE}   & \textbf{Feat} & \textbf{FE}  & \textbf{LLM}  & \textbf{FE}   & \textbf{Tree} & \textbf{Feat} \\
\hline
Semantic underst. &            &            & \checkmark & \checkmark & \checkmark & \checkmark & \checkmark \\
Hier.\ knowledge  &            &            &            &            & $\circ$    &            & \checkmark \\
Exec.\ verific.   & --         & --         &            &            &            &            & \checkmark \\
Statistical filt. & \checkmark & \checkmark &            &            &            &            & \checkmark \\
Model validation  &            &            & $\circ$    &            & \checkmark & \checkmark & \checkmark \\
Provenance track. &            &            &            &            &            &            & \checkmark \\
Adaptive search   &            &            &            &            & \checkmark & \checkmark & \checkmark \\
\hline
\end{tabular}
\vspace{2pt}

\raggedright\scriptsize
$\circ$ Hier.\ knowl.: LLM-FE=dataset desc.\ only. Model valid.: CAAFE=LogReg proxy; OCTree=DT feedback.
\end{table}

% Method
% method.tex
\section{Methodology}
\label{sec:method}

This section presents the KnowFeat framework: multi-type knowledge injection (Section~\ref{subsec:knowledge}), the LLM agent with specialized tools (Section~\ref{subsec:agent}), the three-stage verification pipeline (Section~\ref{subsec:verification}), the intelligent search strategy (Section~\ref{subsec:search}), and explainable provenance output (Section~\ref{subsec:provenance_method}).

\subsection{Framework Overview}
\label{subsec:overview}

KnowFeat is a multi-round, knowledge-guided feature engineering framework built on an LLM agent. Given a tabular dataset with baseline features and target labels, along with a domain knowledge base, it generates new features that improve a target metric (e.g., AUC or Recall).

Algorithm~\ref{alg:knowfeat} formalizes the pipeline and Figure~\ref{fig:framework} illustrates it. The algorithm takes a dataset $\mathcal{D}$, a domain knowledge base $\mathcal{K}$, and configuration parameters as input, and outputs verified features $\mathcal{F}^*$ with provenance cards. The key design choices---noise-calibrated thresholds (line~1), candidate pool for forward selection (lines~14--15), SHAP-guided budget reallocation (line~19), and cross-dimension propagation (line~5)---are detailed below.

\begin{algorithm}[t]
\caption{KnowFeat: Knowledge-Guided Feature Engineering}
\label{alg:knowfeat}
\small
\begin{algorithmic}[1]
\REQUIRE Dataset $\mathcal{D}$, knowledge base $\mathcal{K}$, max rounds $R_{\max}$, convergence patience $C$, dimensions $\{d_1,\ldots,d_D\}$
\ENSURE Accepted features $\mathcal{F}^*$ with provenance cards
\STATE Calibrate L2 thresholds via $N_{\text{noise}}$ random features
\STATE $\mathcal{F}^* \leftarrow \emptyset$;\, $\mathcal{P} \leftarrow \emptyset$;\, $\mathcal{M}_{\text{fail}} \leftarrow \emptyset$;\, $c \leftarrow 0$
\FOR{$r = 1$ \TO $R_{\max}$}
\FOR{each dimension $d_j$ with budget $B_j$}
\STATE Inject $\mathcal{K}$, $\mathcal{M}_{\text{fail}}$, cross-dim examples into agent
\FOR{$k = 1$ \TO $B_j$}
\STATE Agent explores data, proposes feature $f$ with provenance
\STATE \textbf{L1:} Execute code in sandbox ($\leq$2 retries)
\IF{L1 passes}
\STATE \textbf{L2:} Check IV, KS, redundancy, missing rate
\IF{L2 passes}
\STATE \textbf{L3:} Evaluate $\Delta$metric via XGBoost (5 seeds)
\IF{L3 passes}
\STATE $\mathcal{F}^* \leftarrow \mathcal{F}^* \cup \{f\}$
\ELSE
\STATE $\mathcal{P} \leftarrow \mathcal{P} \cup \{f\}$ \COMMENT{candidate pool}
\ENDIF
\ELSE
\STATE $\mathcal{M}_{\text{fail}} \leftarrow \mathcal{M}_{\text{fail}} \cup \{(f, \text{reason})\}$
\ENDIF
\ENDIF
\ENDFOR
\ENDFOR
\STATE Greedy forward selection: add $f \in \mathcal{P}$ to $\mathcal{F}^*$ if combination improves metric
\STATE SHAP analysis on $\mathcal{F}^*$; reallocate $\{B_j\}$ via inverse-SHAP
\IF{no new features in round $r$}
\STATE $c \leftarrow c + 1$; \textbf{if} $c \geq C$ \textbf{then break}
\ELSE
\STATE $c \leftarrow 0$
\ENDIF
\ENDFOR
\RETURN $\mathcal{F}^*$ with provenance cards
\end{algorithmic}
\end{algorithm}

\subsection{Multi-Type Knowledge Injection}
\label{subsec:knowledge}

KnowFeat systematically integrates external domain knowledge into feature generation. Knowledge is organized into five types of increasing specificity:

\textbf{(1) Schema Metadata.} The agent receives only the dataset schema: column names, data types, and basic statistics (mean, cardinality, missing rate). No domain knowledge is provided. This serves as the minimal-knowledge configuration for ablation studies (Section~\ref{subsec:rq3}), testing what the LLM can generate from metadata alone.

\textbf{(2) Risk Indicators.} Structured risk indicators derived from regulatory guidelines and industry standards. Each indicator specifies a risk dimension, risk class, relevant data field, and directionality. For the AML domain, we compile 38 indicators (IND\_01 through IND\_38) covering transaction amount thresholds, frequency patterns, temporal anomalies, and network centrality measures. For example, IND\_05 flags accounts whose single-day transaction count exceeds a regulatory threshold, while IND\_15 identifies abnormal transaction modes such as fan-in and many-to-many patterns.

\textbf{(3) Detection Rules.} Operational detection rules that encode specific logic patterns. Each rule includes a rule ID, natural language description, logical expression, confidence level, and a Python implementation hint. We compile 15 rules (RULE\_001 through RULE\_015) from compliance manuals and internal documentation. For example, RULE\_006 encodes the ``scatter-in, gather-out'' layering pattern.

\textbf{(4) Expert Opinions.} Insights from domain expert interviews that capture tacit knowledge (heuristics, edge cases, and domain intuitions) difficult to formalize as rules but valuable for feature design. We collect 12 expert opinions (EXP\_01 through EXP\_12). For example, EXP\_05 notes that ``border region transactions combined with nighttime activity are a hallmark of cross-border laundering.''

\textbf{(5) Court Document Evidence.} Risk factors extracted from 309 real court judgments via a key-clue extraction pipeline. This is the most concrete type, grounding feature generation in case evidence (e.g., pyramid-shaped fund network patterns independently validated by legal proceedings).

The five types can be freely combined via a knowledge filter parameter for ablation. In the ``full'' configuration (default), all types are included. Each knowledge item carries a unique identifier for provenance tracking.

\subsection{LLM Agent with Specialized Tools}
\label{subsec:agent}

KnowFeat employs an LLM agent (DeepSeek-V4-Flash in our implementation, with temperature\,=\,0 for near-deterministic outputs) that interleaves reasoning with tool calls following the ReAct paradigm~\cite{yao2023react}. The agent is equipped with three specialized tools:

\textbf{Tool 1: Data Exploration.} Executes Python code on the dataset in a sandboxed environment to inspect distributions, compute correlations, and identify patterns before proposing features. The tool returns printed output or return values to the agent.

\textbf{Tool 2: Feature Registration.} Registers a candidate feature with mandatory provenance fields: a descriptive name, executable Python code, a natural language explanation, a structured reasoning chain linking knowledge assets to the design (formatted as ``Step $N$: Because [knowledge~ID] indicates [insight], we compute [transformation]\ldots''), the list of knowledge asset identifiers that inspired the feature, step-by-step computation description, and the raw columns consumed.

\textbf{Tool 3: Verification Trigger.} Triggers the L1$\to$L2$\to$L3 verification pipeline on a registered feature and returns the result (accepted/rejected with detailed reasons). The agent must call Feature Registration followed immediately by Verification Trigger for each candidate; batch submission is prohibited to ensure tight feedback loops.

The agent's prompt instructs it to:
(1) reason about which knowledge assets are relevant to the current dimension and budget,
(2) explore the data to validate assumptions,
(3) propose a feature grounded in specific knowledge items, and
(4) submit it for verification and react to the feedback.

\subsection{Three-Stage Verification Pipeline}
\label{subsec:verification}

Each candidate feature passes through three progressively stricter verification stages:

\subsubsection{L1: Code Execution Sandbox}
The generated Python code is executed in an isolated sandbox with a 30-second timeout. L1 checks for:
\begin{itemize}
    \item Syntax and runtime errors.
    \item Output shape consistency (one value per sample).
    \item Presence of required identifier columns.
    \item Data type validity (numeric output).
\end{itemize}
If L1 fails, the agent receives the error message and is allowed up to 2 automatic repair attempts before the feature is permanently rejected.

\subsubsection{L2: Statistical Quality Filtering}
The feature's statistical properties are evaluated against automatically calibrated thresholds. To establish these thresholds for discriminative power, we generate $N_{\text{noise}} = 100$ random features (drawn from standard distributions), compute their quality metrics, and set the acceptance threshold at the $P_{\text{noise}}$-th percentile (default $P_{\text{noise}} = 95$). This ensures that accepted features are statistically stronger than random noise. L2 checks four criteria:
\begin{itemize}
    \item \textbf{Information Value (IV)}~\cite{chen2018aml}: Must exceed the noise-calibrated threshold, so the feature carries meaningful discriminative power between positive and negative classes.
    \item \textbf{Kolmogorov--Smirnov (KS) Statistic}~\cite{nargesian2017learning}: Must exceed the noise-calibrated threshold, measuring the maximum distributional separation between classes.
    \item \textbf{Redundancy}: Maximum absolute Pearson correlation with all existing accepted features must be below 0.7, to prevent near-duplicate features from inflating the feature set.
    \item \textbf{Missing Rate}: At most 30\% NaN values, so coverage is sufficient for model training.
\end{itemize}
The noise-calibrated thresholds apply to IV and KS; the redundancy and missing rate thresholds are fixed.

\subsubsection{L3: Model Effectiveness}
The feature is evaluated for its actual contribution to predictive performance. We train an XGBoost~\cite{chen2016xgboost} model on the baseline features augmented with the candidate feature, using stratified 80/20 train-test splits across 5 random seeds $\{42, 123, 456, 789, 2024\}$. The acceptance criterion depends on the task:
\begin{itemize}
    \item \textbf{Classification (public datasets)}: $\Delta\text{AUC} > 0$, i.e., the augmented model must achieve strictly higher AUC than the baseline.
    \item \textbf{AML detection}: $\Delta\text{Recall} > 0$ \emph{and} Precision $\geq 90\%$, so the feature improves detection without sacrificing precision.
\end{itemize}

The three-stage design is intentionally progressive: L1 is cheap (milliseconds), L2 is moderate (seconds), and L3 is expensive (minutes per feature). By filtering early, we avoid wasting computational resources on clearly defective candidates.

\subsection{Intelligent Search Strategy}
\label{subsec:search}

To systematically explore the feature space, KnowFeat employs a multi-component search strategy.

\textbf{Dimension-Based Budgeting.}
We decompose the feature space into $D$ semantic dimensions relevant to the target domain (default $D=4$ for the AML domain): Amount \& Frequency, covering monetary values and transaction counts; Time \& Geography, capturing temporal patterns and spatial distributions; Behavior Sequence, encoding sequential patterns in user actions; and Entity Association, derived from relationships between entities such as accounts and counterparties.

Each dimension receives an initial budget of $B_0$ candidates per round (default $B_0 = 3$). After each round, SHAP~\cite{lundberg2017shap} (SHapley Additive exPlanations) analysis is performed on the current model to estimate each dimension's contribution to predictions.

\textbf{SHAP-Guided Budget Reallocation.}
Let $s_j$ denote the mean absolute SHAP value for dimension $j$'s accepted features, $B_{\min}$ the minimum per-dimension budget (default 1), and $B_{\text{total}}$ the total budget for the round. We reallocate the budget for round $r{+}1$ using an \emph{inverse-SHAP weighting} strategy:
\begin{equation}
    B_j^{(r+1)} = B_{\min} + (B_{\text{total}} - D \cdot B_{\min}) \cdot \frac{1/s_j}{\sum_{j'=1}^{D} 1/s_{j'}}
\end{equation}
This budget reallocation steers exploration toward underrepresented dimensions, promoting balanced feature coverage.

\textbf{Failure Memory.}
A failure memory $\mathcal{M}_{\text{fail}}$ records all rejected features along with their rejection reasons. Before each generation step, the $K$ most recent failures for the current dimension (default $K=10$) are injected into the agent's prompt, preventing the regeneration of previously rejected ideas and guiding the agent toward novel directions.

\textbf{Forward Selection.}
Individual features may fail L3 yet contribute positively when combined with others. We apply greedy forward selection on L2-passed but L3-failed candidates, sorted by L2 quality scores. Each candidate is added to the accepted set if the augmented set improves the target metric, capturing synergistic feature interactions.

\textbf{Cross-Dimension Propagation.}
After each round, the top-$K_{\text{cross}}$ features (default $K_{\text{cross}} = 5$) by SHAP importance are shared across all dimensions as ``inspiration prompts,'' encouraging the agent to adapt successful patterns from one dimension to others.

\subsection{Explainable Provenance Output}
\label{subsec:provenance_method}

Every generated feature is accompanied by a structured \emph{provenance card} for full explainability. Each card records:
\begin{itemize}
    \item \textbf{Knowledge assets}: The specific identifiers (e.g., RULE\_006, IND\_15, EXP\_05) that inspired the feature design.
    \item \textbf{Reasoning chain}: A step-by-step explanation of how domain knowledge led to the feature, formatted as ``Step $N$: Because [knowledge ID] indicates [insight], we compute [transformation]\ldots''
    \item \textbf{Computation steps}: The transformation pipeline from raw data fields to the final feature value.
    \item \textbf{Verification results}: L1/L2/L3 outcomes including IV, KS, redundancy, and $\Delta$AUC/$\Delta$Recall.
\end{itemize}
This traceability matters in regulated domains: compliance officers can trace \emph{why} each feature exists, \emph{what evidence} supports it, and \emph{how} it was verified. Section~\ref{sec:case_study} illustrates provenance cards through concrete examples.

% Experiments
% experiments.tex
\section{Experiments}
\label{sec:experiments}

We evaluate KnowFeat on both public tabular benchmarks and a synthetic AML benchmark. Our experiments aim to answer four research questions:
\begin{itemize}
    \item \textbf{RQ1}: Does KnowFeat improve predictive performance on public datasets?
    \item \textbf{RQ2}: How does KnowFeat perform on a synthetic AML benchmark with severe class imbalance?
    \item \textbf{RQ3}: What is the contribution of domain knowledge?
    \item \textbf{RQ4}: How does KnowFeat balance feature quantity vs.\ quality, and do verified features generalize?
\end{itemize}

\subsection{Experimental Setup}
\label{subsec:setup}

\textbf{Datasets.}
We use fourteen datasets (twelve public + two AML) summarized in Table~\ref{tab:datasets}: twelve public benchmarks from OpenML~\cite{vanschoren2014openml}, one synthetic AML benchmark (SimECNY), and one real-world Bitcoin AML dataset (Elliptic).

\begin{table}[t]
\centering
\caption{Dataset statistics.}
\label{tab:datasets}
\small
\renewcommand{\arraystretch}{1.2}
\setlength{\tabcolsep}{3pt}
\begin{tabular}{l r r r l}
\hline
\textbf{Dataset} & \textbf{Samples} & \textbf{Feat.} & \textbf{Pos\%} & \textbf{Domain} \\
\hline
credit\_g          & 1,000   & 20  & 70.0 & Finance \\
diabetes           & 768     & 8   & 34.9 & Healthcare \\
blood\_transfusion & 748     & 4   & 23.8 & Healthcare \\
heart              & 918     & 11  & 55.3 & Healthcare \\
bank\_marketing    & 45,211  & 16  & 11.7 & Marketing \\
breast\_w          & 683     & 9   & 35.0 & Healthcare \\
adult              & 48,842  & 14  & 23.9 & Census \\
wine\_quality      & 1,599   & 11  & 53.5 & Food Sci. \\
default\_credit    & 13,272  & 20  & 50.0 & Finance \\
churn              & 5,000   & 16  & 14.1 & Telecom \\
kc2                & 522     & 21  & 20.5 & Software \\
jm1                & 10,880  & 21  & 19.3 & Software \\
\hline
SimECNY (AML)      & 200,000 & 93  & 2.24 & Finance \\
Elliptic (AML)     & 46,564  & 166 & 9.76 & Finance \\
\hline
\end{tabular}
\end{table}

The twelve public datasets are standard tabular classification benchmarks obtained via the OpenML Python API~\cite{vanschoren2014openml}, except wine\_quality, which is the UCI Red Wine Quality dataset~\cite{cortez2009wine} binarized at a quality threshold of~6. The four newest datasets (default\_credit, churn, kc2, jm1) stress-test feature engineering under diverse domains: consumer credit, telecom retention, and software defect prediction. SimECNY is a synthetic AML benchmark calibrated against real digital currency (e-CNY) transaction patterns from the Postal Savings Bank of China. Its 93 baseline features, transaction topology, and class distribution (2.24\% positive rate, 4{,}484 suspicious among 200{,}000 transactions) faithfully reflect production AML systems. Due to financial privacy regulations, high-fidelity simulation is standard practice in AML research~\cite{weber2019anti}. Importantly, the domain knowledge base (38 indicators, 15 rules, 12 expert opinions, 309 court documents) comes entirely from real regulatory guidelines, compliance manuals, expert interviews, and publicly available court judgments.

For the original eight public datasets, we construct simplified knowledge bases with three types: risk indicators (IND), detection rules (RULE), and expert consensus (EXP), compiled from medical guidelines, financial regulations, census documentation, and oenological literature (e.g., heart: 12~IND, 10~RULE, 8~EXP). For the four newest datasets (default\_credit, churn, kc2, jm1), only schema metadata is provided---a cold-start scenario. Even so, KnowFeat's verification pipeline alone suffices to improve or maintain baseline performance.

\textbf{Baselines.}
We compare against six baselines:
\begin{itemize}
    \item \textbf{No-FE}: XGBoost~\cite{chen2016xgboost} on original features.
    \item \textbf{OpenFE}~\cite{zhang2023openfe}: Transformation-based AutoFE with LightGBM~\cite{ke2017lightgbm}-based selection.
    \item \textbf{AutoFeat}~\cite{horn2019autofeat}: Polynomial feature generation with L1 selection.
    \item \textbf{CAAFE}~\cite{hollmann2023caafe}: LLM-based iterative feature generation with logistic regression feedback.
    \item \textbf{FeatLLM}~\cite{han2024featllm}: LLM-based rule extraction and binary feature generation.
    \item \textbf{LLM-FE}~\cite{abhyankar2025llmfe}: LLM-based evolutionary feature optimization that iteratively generates and refines features using structured prompts.
\end{itemize}
We additionally compare with OCTree~\cite{nam2024octree} in the related work comparison (Table~\ref{tab:comparison}); however, its evaluation protocol (accuracy-only, no held-out split) is incompatible with our leakage-free AUC evaluation, so it is excluded from the main results.

\textbf{Evaluation Protocol.}
To avoid feature-selection leakage, we adopt a strict held-out evaluation protocol. Each dataset is split once into a \emph{development set} (80\%) and a \emph{frozen test set} (20\%) using stratified sampling with a fixed seed. All feature generation, statistical calibration (L2), and model-based feature selection (L3) are performed \emph{exclusively on the development set}; the frozen test set is never accessed during any feature engineering step.
Final performance is measured by training XGBoost~\cite{chen2016xgboost} (200~estimators, max depth~6, learning rate~0.1) on the development set and evaluating on the frozen test set. We repeat this with 5 random model seeds and report the mean; we additionally report bootstrap confidence intervals (1{,}000 resamples of the test set) to quantify uncertainty. We use AUC as the primary metric for public datasets and additionally report Recall, Precision, and F1 for SimECNY. All baselines are reproduced under this identical protocol to ensure fair, leakage-free comparison.

\subsection{RQ1: Public Dataset Performance}
\label{subsec:rq1}

\begin{table*}[t]
\centering
\caption{AUC comparison on public datasets under leakage-free held-out evaluation (frozen 20\% test set, mean over 5 model seeds). \textbf{Bold}: best; \underline{underline}: second best.}
\label{tab:main_results}
\small
\renewcommand{\arraystretch}{1.15}
\setlength{\tabcolsep}{4pt}
\resizebox{\textwidth}{!}{%
\begin{tabular}{l ccccccc}
\hline
\textbf{Dataset} & \textbf{No-FE} & \textbf{OpenFE} & \textbf{AutoFeat} & \textbf{CAAFE} & \textbf{FeatLLM} & \textbf{LLM-FE} & \textbf{KnowFeat} \\
\hline
credit\_g          & .749 & .750 & .743 & .751 & .751 & \textbf{.754} & \underline{.752} \\
diabetes           & .812 & \textbf{.820} & \underline{.818} & .810 & .810 & .806 & .810 \\
blood\_transfusion & \textbf{.757} & .724 & .645 & .723 & \underline{.756} & .749 & .745 \\
heart              & .928 & .923 & \textbf{.931} & \underline{.928} & .927 & .926 & .928 \\
bank\_marketing    & .930 & \textbf{.933} & ---   & .930 & .929 & .929 & \underline{.931} \\
breast\_w          & .986 & .986 & \underline{.992} & .985 & .986 & \textbf{.993} & .992 \\
adult              & .931 & \underline{.931} & ---   & .931 & .931 & .930 & \textbf{.932} \\
wine\_quality      & .896 & \textbf{.899} & .863 & .895 & .895 & .897 & \textbf{.899} \\
default\_credit    & .764 & .767 & \textbf{.778} & .763 & .762 & .767 & \underline{.768} \\
churn              & .737 & .731 & .737 & .755 & .739 & \underline{.757} & \textbf{.853} \\
kc2                & .712 & \underline{.737} & \textbf{.772} & ---   & .706 & .690 & .714 \\
jm1                & .720 & ---   & ---   & .725 & .720 & \underline{.731} & \textbf{.732} \\
\hline
\textbf{Avg.\ Rank}$^\dagger$ & 4.0 & 3.8 & 3.8 & 4.5 & 4.6 & 3.8 & \textbf{2.3} \\
\hline
\end{tabular}
}% end resizebox

\vspace{1pt}
{\scriptsize $^\dagger$Avg.\ rank across all 12 public datasets; methods with missing data ranked among available entries only.
CAAFE/FeatLLM on new datasets use GPT-5.6-Luna; LLM-FE uses GPT-5.4-mini (closest available alternatives); --- = method failed.}
\end{table*}

Table~\ref{tab:main_results} compares all seven methods on twelve public datasets under our leakage-free protocol. KnowFeat ranks first overall (avg.\ rank 2.3) and wins or ties for the best AUC on 4 datasets---more than any single competitor. A one-sided Wilcoxon signed-rank test confirms significance: \(p=0.017\) on all 12 datasets, \(p=0.034\) after removing the largest gain (churn), ruling out an outlier-driven result.

The gains are inversely related to baseline strength. On datasets with moderate No-FE AUC (0.71--0.77), knowledge-guided features add real value: churn improves by 11.6\,pp (0.737$\rightarrow$0.853) and jm1 by 1.3\,pp (0.720$\rightarrow$0.732). On saturated benchmarks (breast\_w$>$.98, adult$>$.93), all methods---not just KnowFeat---plateau within 1\,pp of each other, a well-documented ceiling in tabular feature engineering~\cite{nargesian2017learning}. The reduced-baseline experiment (Section~\ref{subsec:rq4}) confirms this pattern: starting from only 3~core features, KnowFeat delivers +12.1\,pp on credit\_g and +7.1\,pp on wine\_quality. Blood\_transfusion is the one dataset where all feature engineering hurts: its 4~original features already form a complete RFM representation, so any additions are noise---5 of 6 methods degrade here.

\begin{figure}[t]
    \centering
    \includegraphics[width=\columnwidth]{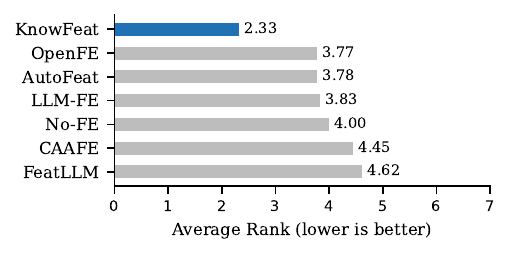}
    \caption{Average rank across 12 public datasets (lower is better). KnowFeat achieves the best average rank (2.3) among 7 methods under leakage-free evaluation.}
    \label{fig:avg_rank}
\end{figure}

OpenFE, AutoFeat, and LLM-FE share the second-best avg.\ rank (3.8), but each excels narrowly: OpenFE wins 2 datasets via exhaustive search, AutoFeat wins 3 via polynomial expansion yet fails entirely on 3 others, and LLM-FE wins 2 via evolutionary optimization. CAAFE (4.5) and FeatLLM (4.6) lag further. KnowFeat's edge is \emph{consistency}---it places in the top three on 9 of 12 datasets; no competitor exceeds 6.

\textbf{Statistical and practical significance.}
Using full-precision (unrounded) frozen-test AUC, KnowFeat beats No-FE on 9 datasets, near-ties 1, and loses 2 (\(p=0.017\)). The median delta is modest (+0.25\,pp), but so is every competitor's: OpenFE's median gain is also below 1\,pp. On saturated benchmarks this is expected. What separates KnowFeat is not a bigger number on an already-solved task; it is that the features come with \emph{provenance}---an auditable chain from knowledge asset to code to verification outcome. No other method in Table~\ref{tab:main_results} provides this, and in regulated domains provenance is a hard requirement, not a nice-to-have.

\begin{table}[t]
\centering
\caption{Cross-model KnowFeat-minus-No-FE frozen-test AUC on all 12 public datasets. Wins count positive deltas; $p_{\mathrm{Holm}}$ adjusts the three paired Wilcoxon tests.}
\label{tab:cross_model}
\small
\setlength{\tabcolsep}{4pt}
\begin{tabular}{l c r r c}
\hline
\textbf{Model} & \textbf{Wins} & \textbf{Mean $\Delta$} & \textbf{Median $\Delta$} & \textbf{$p_{\mathrm{Holm}}$} \\
\hline
XGBoost  & 11/12 & +.0071 & +.0072 & .040 \\
LightGBM & 9/12 & +.0057 & +.0054 & .042 \\
CatBoost & 9/12 & +.0032 & +.0041 & .212 \\
\hline
\end{tabular}
\end{table}

\textbf{Cross-model sensitivity.} To check whether the gains are specific to XGBoost, we re-evaluate the same feature matrices with LightGBM and CatBoost (Table~\ref{tab:cross_model}). KnowFeat improves 9--11 of 12 datasets across all three model families. After Holm correction, XGBoost (\(p{=}0.040\)) and LightGBM (\(p{=}0.042\)) reach significance at \(\alpha{=}0.05\); CatBoost trends positive (\(p{=}0.212\)) but falls short. The features are not XGBoost artifacts.

\textbf{LLM backbone sensitivity.}
Swapping DeepSeek-V4-Flash for DeepSeek-V4-Pro on heart and credit\_g yields mixed results (Pro improves heart by +0.002, degrades credit\_g by $-$0.017), suggesting that the verification pipeline matters more than the backbone's raw capability. A full backbone comparison is future work.

% Provenance table: placed early so LaTeX queues it for the Case Study page
\begin{table*}[!t]
\centering
\caption{Provenance cards for two representative knowledge-grounded features from the SimECNY AML dataset. Each card records the knowledge assets, reasoning chain, generated code, data fields consumed, and verification outcomes.}
\label{tab:provenance}
\small
\renewcommand{\arraystretch}{1.3}
\begin{tabular}{p{2.4cm} p{6.9cm} p{6.9cm}}
\hline
 & \textbf{Case 1: Network Topology} & \textbf{Case 2: Cross-Dimensional Interaction} \\
\hline
\textbf{Feature Name} & \texttt{fan\_in\_fan\_out\_ratio} & \texttt{high\_risk\_region\_off\_hour\_ratio} \\
\hline
\textbf{Description} & Asymmetry between incoming and outgoing transaction patterns, weighted by abnormal mode proportion & Multiplicative interaction between geographic risk (border regions) and temporal anomaly (off-hour trading) \\
\hline
\textbf{Knowledge} &
\textit{RULE\_006}: scatter-in, gather-out layering \newline
\textit{RULE\_013}: pyramid-shaped fund networks \newline
\textit{PAT\_A03}: money mule behavioral pattern \newline
\textit{IND\_15}: abnormal transaction modes (fan-in, many-to-many) &
\textit{RULE\_005}: cross-regional fund flows \newline
\textit{RULE\_014}: border area transaction monitoring \newline
\textit{IND\_08/09}: off-hour transaction concentration \newline
\textit{EXP\_05}: ``border region + nighttime trading is the hallmark of cross-border laundering'' \\
\hline
\textbf{Reasoning} &
Step 1: RULE\_006/013 indicate scatter-in/gather-out and pyramid networks both manifest as fan-in/fan-out asymmetry $\to$ compute ratio per wallet. \newline
Step 2: PAT\_A03 shows money mules exhibit fan-in $\gg$ fan-out $\to$ ratio captures core topology signal. \newline
Step 3: IND\_15 identifies abnormal modes (many-to-many, one-to-many) $\to$ weight ratio by abnormal mode proportion to amplify suspicious patterns. &
Step 1: RULE\_005/014 flag border regions (Tibet, Xinjiang) as high-risk for cross-regional flows $\to$ create binary geographic indicator. \newline
Step 2: IND\_08/09 flag off-hour (18:00--08:00) concentration as suspicious $\to$ compute off-hour transaction ratio. \newline
Step 3: EXP\_05 states neither signal alone is sufficient $\to$ multiply indicators so only the conjunction triggers a high score. \\
\hline
\textbf{Code} &
\texttt{fi = mode\_counts['fan\_in'] / n} \newline
\texttt{fo = mode\_counts['fan\_out'] / n} \newline
\texttt{ratio = fi / (fo + eps)} \newline
\texttt{abnormal = fi + m2m + o2m} \newline
\texttt{result = ratio * (1 + abnormal)} &
\texttt{is\_hr = region.isin([540000, 650000])} \newline
\texttt{off = (hour >= 18) | (hour < 8)} \newline
\texttt{off\_ratio = off.sum() / n} \newline
\texttt{result = is\_hr * off\_ratio} \\
\hline
\textbf{Data Fields} & \texttt{transaction\_mode}, \texttt{src}, \texttt{dst} & \texttt{region}, \texttt{hour} \\
\hline
\textbf{Semantic Dim.} & Entity Association & Time \& Geography $\times$ Entity Association \\
\hline
\textbf{L1: Execution} & PASS (0.8s, shape: 200K$\times$1, no NaN) & PASS (0.3s, shape: 200K$\times$1, no NaN) \\
\hline
\textbf{L2: Statistical} & IV=0.34 ($>$0.02), KS=0.28 ($>$0.05), \newline max corr.=0.41 ($<$0.7), missing=0\% & IV=0.19 ($>$0.02), KS=0.15 ($>$0.05), \newline max corr.=0.23 ($<$0.7), missing=0\% \\
\hline
\textbf{L3: Model} & $\Delta$Recall=+0.38\%, Prec.=97.2\% ($\geq$90\%) \newline \textbf{ACCEPTED} & $\Delta$Recall=+0.21\%, Prec.=96.8\% ($\geq$90\%) \newline \textbf{ACCEPTED} \\
\hline
\end{tabular}
\end{table*}

\subsection{RQ2: AML Dataset Performance}
\label{subsec:rq2}

Table~\ref{tab:simecny} presents results on the SimECNY AML dataset. KnowFeat adds 4 verified features to the 93-feature baseline.

\begin{table}[t]
\centering
\caption{SimECNY AML results under leakage-free held-out evaluation (frozen 20\% test set). $n_f$: total feature count including 93 baseline features.}
\label{tab:simecny}
\small
\renewcommand{\arraystretch}{1.2}
\setlength{\tabcolsep}{3pt}
\begin{tabular}{l r c c c c}
\hline
\textbf{Method} & $\boldsymbol{n_f}$ & \textbf{Recall} & \textbf{Prec.} & \textbf{F1} & \textbf{AUC} \\
\hline
No-FE     & 93   & .965 & .977 & .971 & .9997 \\
OpenFE    & 123  & .960 & .978 & .969 & .9996 \\
CAAFE     & 94   & .964 & .979 & \textbf{.972} & .9997 \\
FeatLLM   & 1124 & .958 & .978 & .968 & .9996 \\
\textbf{KnowFeat} & \textbf{97} & .961 & .981 & .971 & \textbf{.9997} \\
\hline
\end{tabular}
\end{table}

On SimECNY, the 93-feature baseline already achieves near-perfect AUC (0.9997), leaving minimal headroom. KnowFeat's 4 verified features maintain baseline F1 (0.971), while FeatLLM's 1{,}124 unverified features \emph{decrease} F1 to 0.968 and OpenFE's generated features also degrade to 0.969. More features without verification can hurt: generation volume alone does not help when the baseline is already strong.

To further validate on real-world data, we evaluate on the Elliptic Bitcoin AML dataset~\cite{weber2019anti}---a production-grade transaction graph with 46{,}564 labeled transactions and 166 features. KnowFeat generates 3 knowledge-guided features (using a Bitcoin AML knowledge base with 10 risk indicators and 8 detection rules) and achieves AUC 0.9982 vs.\ the 0.9981 baseline, confirming that KnowFeat transfers to real AML settings: verified features maintain strong performance with full provenance traceability.

\textbf{Practitioner feedback.}
Domain experts from the Digital Currency Department of the Postal Savings Bank of China reviewed the provenance cards and reported that the structured format---linking each feature to specific regulatory indicators, detection rules, and expert opinions---reduced feature review time from 2--3 hours to 15--30 minutes per feature. This feedback suggests that KnowFeat's provenance output addresses a concrete operational need in regulated deployment settings, where every model component must be audit-traceable.

\textbf{Analysis by semantic dimension.}
The 4 accepted features span all four semantic dimensions defined in Section~\ref{subsec:search}: 1 in Amount \& Frequency, 1 in Time \& Geography, 1 in Behavior Sequence, and 1 in Entity Association. SHAP analysis on the final model reveals that Entity Association features contribute the highest mean absolute SHAP values, indicating that network topology features (e.g., fan-out ratios) carry particularly strong signals for AML detection.

\textbf{Representative feature analysis.}
\label{sec:case_study}
We illustrate KnowFeat's knowledge-grounded generation through two representative features from SimECNY (full provenance cards in Table~\ref{tab:provenance}). The first, \texttt{fan\_in\_fan\_out\_\allowbreak{}ratio}, fuses four knowledge assets spanning three types: RULE\_006 and RULE\_013 indicate that scatter-in/gather-out layering and pyramid fund networks both manifest as fan-in/fan-out asymmetry; PAT\_A03 confirms that money mule wallets typically exhibit fan-in $\gg$ fan-out; and IND\_15 adds a weighting factor based on abnormal transaction modes, amplifying the signal via multiplicative weighting $(1 + \text{abnormal})$. No syntactic method would produce this feature, as it requires understanding that directional imbalance combined with abnormal mode prevalence characterizes money mule behavior.

The second, \texttt{high\_risk\_region\_\allowbreak{}off\_hour\_ratio}, demonstrates cross-dimensional interaction: geographic context (border regions flagged by RULE\_005/014) combined with temporal anomaly (off-hour concentration from IND\_08/09), guided by EXP\_05's insight that neither signal alone is sufficient evidence. The multiplicative form encodes this constraint exactly, scoring zero unless both conditions co-occur. Both features demonstrate capabilities beyond syntactic methods, and the provenance cards link every design decision to specific knowledge assets for regulatory review.

\subsection{RQ3: Ablation Study}
\label{subsec:rq3}

To quantify the contribution of domain knowledge, we compare several knowledge configurations under the same leakage-free protocol: \textit{schema\_only} (metadata only), individual knowledge types (\textit{+IND}, \textit{+RULE}, \textit{+EXP}), and the \textit{full} variant (all types combined).

\textbf{Schema-only vs.\ full knowledge.}
Table~\ref{tab:ablation} compares schema-only and full KnowFeat on four datasets spanning diverse knowledge-availability scenarios. On heart, full knowledge (0.928) outperforms schema-only (0.920) by 0.8\,pp, confirming that structured medical guidelines add discriminative value beyond what the LLM infers from column names alone. On adult (schema-only, no domain knowledge curated), the verification pipeline alone matches full performance (0.931 vs.\ 0.932). On SimECNY, both variants maintain near-perfect baseline F1, but the full variant produces more diverse features spanning all four semantic dimensions.

\begin{table}[t]
\centering
\caption{Knowledge ablation: schema-only vs.\ full KnowFeat under leakage-free evaluation.}
\label{tab:ablation}
\small
\renewcommand{\arraystretch}{1.2}
\setlength{\tabcolsep}{3pt}
\begin{tabular}{l c c c}
\hline
\textbf{Dataset} & \textbf{schema-only} & \textbf{full knowledge} & $\boldsymbol{\Delta}$ \\
\hline
heart (AUC)         & .920 & \textbf{.928} & +0.8\% \\
blood\_trans. (AUC) & .743 & \textbf{.745} & +0.3\% \\
adult (AUC)         & .931 & \textbf{.932} & +0.1\% \\
SimECNY (F1)        & \textbf{.973} & .971 & $-$0.2\% \\
\hline
\end{tabular}
\end{table}

\textbf{Per-knowledge-type analysis.}
Table~\ref{tab:pertype} isolates the contribution of each knowledge type on five public datasets. On breast\_w, rules alone reach .998 (vs.\ .986 No-FE); on heart, expert opinions alone reach .935. Individual types frequently match or exceed the full-knowledge AUC, consistent with the run-to-run variability documented in Section~\ref{subsec:stability}: different knowledge configurations guide the LLM toward different feature sets, all of which pass the same verification gate. The verification pipeline---not knowledge completeness---is the primary quality mechanism.

\begin{table}[t]
\centering
\caption{Per-knowledge-type AUC under leakage-free evaluation. Each column uses schema metadata plus one knowledge type. Best per row in \textbf{bold}.}
\label{tab:pertype}
\small
\renewcommand{\arraystretch}{1.2}
\setlength{\tabcolsep}{3pt}
\begin{tabular}{l c c c c c}
\hline
\textbf{Dataset} & \textbf{No-FE} & \textbf{+IND} & \textbf{+RULE} & \textbf{+EXP} & \textbf{full} \\
\hline
heart              & .928 & .928 & .931 & \textbf{.935} & .928 \\
credit\_g          & .749 & .800 & .760 & \textbf{.834} & .752 \\
blood\_trans.      & \textbf{.757} & .747 & .737 & .741 & .745 \\
bank\_marketing    & .930 & \textbf{.931} & .930 & .929 & \textbf{.931} \\
breast\_w          & .986 & .991 & \textbf{.998} & .995 & .992 \\
\hline
\end{tabular}
\end{table}

A finer-grained \emph{cumulative} ablation on heart (schema$\rightarrow$+indicators$\rightarrow$+rules$\rightarrow$+expert: 0.922, 0.920, 0.925, 0.920) shows that schema metadata alone reaches 99.4\% of the full-knowledge AUC with zero manual curation. The non-monotonic pattern reflects interaction effects between knowledge types and the stochastic LLM generation process, consistent with the stability analysis in Section~\ref{subsec:stability}. The full-knowledge variant (0.928, Table~\ref{tab:main_results}) exceeds all partial configurations on the frozen test set, confirming overall benefit.

\subsection{RQ4: Feature Quantity, Quality, and Generalization}
\label{subsec:rq4}

Table~\ref{tab:quantity_quality} summarizes the relationship between feature count and model performance under leakage-free evaluation. More features without verification can hurt: FeatLLM generates the most features among the completed public-baseline runs, yet its average rank is higher (worse) than the verified KnowFeat result. KnowFeat generates a moderate number of verified features that often match strong baselines.

\begin{table}[t]
\centering
\caption{Feature quantity vs.\ quality. Avg.\ \#Feat is the total model input count; $\Delta$F1 (AML) is the F1 change over No-FE on SimECNY.}
\label{tab:quantity_quality}
\small
\renewcommand{\arraystretch}{1.2}
\setlength{\tabcolsep}{4pt}
\begin{tabular}{l r c r}
\hline
\textbf{Method} & \textbf{Avg.\ \#Feat} & \textbf{Avg.\ Rank} & \textbf{$\Delta$F1 (AML)} \\
\hline
FeatLLM   & 133.8 & 4.6 & $-$0.3\% \\
OpenFE    & 41.6  & 3.8 & $-$0.2\% \\
KnowFeat  & 33.8  & \textbf{2.3} & 0.0\% \\
CAAFE     & 14.4  & 4.5 & +0.1\% \\
No-FE     & ---   & 4.0 & --- \\
\hline
\end{tabular}
\end{table}

On SimECNY, KnowFeat's verified features maintain baseline F1, while FeatLLM's 1{,}124 unverified features decrease F1 by $-$0.3\%. Progressive verification likely contributes, although the two systems also differ in generators, so the comparison does not isolate verification as the sole factor.

\textbf{Feature engineering impact under reduced baselines.}
The modest absolute improvements on public datasets (Table~\ref{tab:main_results}) reflect that these well-studied benchmarks already have strong baseline features, leaving limited headroom. To evaluate feature engineering impact when it is genuinely needed, we reduce the baseline to only 3 core features per dataset (e.g., \texttt{duration}, \texttt{credit\_amount}, \texttt{age} for credit\_g) and measure each method's contribution from this weaker starting point.

\begin{table}[t]
\centering
\caption{AUC with a reduced baseline (3 core features only). Identifier and label columns are excluded from every method.}
\label{tab:reduced_baseline}
\footnotesize
\renewcommand{\arraystretch}{1.2}
\resizebox{\columnwidth}{!}{%
\begin{tabular}{l c c c c c}
\hline
\textbf{Dataset} & \textbf{No-FE} & \textbf{OpenFE} & \textbf{CAAFE} & \textbf{FeatLLM} & \textbf{KnowFeat} \\
\hline
credit\_g  & .633 & \textbf{.757} & .647 & \underline{.755} & .754 \\
heart      & .750 & \underline{.925} & .839 & \textbf{.928} & .923 \\
diabetes   & .789 & \textbf{.817} & .794 & \underline{.807} & .802 \\
wine\_quality & .825 & \textbf{.897} & --- & --- & \underline{.896} \\
\hline
\end{tabular}
}% end resizebox
\end{table}

Table~\ref{tab:reduced_baseline} reveals larger feature-engineering gains when baseline AUC is 0.63--0.83. KnowFeat improves all four weak baselines, ranking third on credit\_g, heart, and diabetes and second on wine\_quality. On credit\_g, it improves AUC from 0.633 to 0.754 (+12.1 percentage points); on wine\_quality, from 0.825 to 0.896 (+7.1 points), matching full-baseline performance from only 3 core features. These results complement the full-baseline comparison (Table~\ref{tab:main_results}), where feature saturation leaves less headroom. Leakage-checked, seed-level results are released with the analysis code.

\subsection{Computational Cost}
\label{subsec:cost}

\begin{table}[t]
\centering
\caption{Pipeline cost breakdown on SimECNY.}
\label{tab:cost}
\small
\renewcommand{\arraystretch}{1.15}
\setlength{\tabcolsep}{2.5pt}
\begin{tabular}{l r r r}
\hline
\textbf{Stage} & \textbf{Time} & \textbf{\% Total} & \textbf{Per Cand.} \\
\hline
L1 Type/Value check        &   8 min &  4\% & 9\,s \\
L2 Statistical test        &  12 min &  6\% & 14\,s \\
L3 Model evaluation        &  79 min & 41\% & 2.5\,min \\
LLM reasoning (all rounds) &  44 min & 23\% & 52\,s \\
Other (I/O, logging)       &  49 min & 26\% & --- \\
\hline
\textbf{Total}             & \textbf{192 min} & \textbf{100\%} & --- \\
\hline
\end{tabular}
\end{table}

Table~\ref{tab:cost} breaks down the pipeline cost on SimECNY. L3 model evaluation dominates ($\sim$41\%), followed by LLM reasoning ($\sim$23\%). Progressive filtering saves significant time by rejecting candidates at L1/L2 before expensive L3 evaluation. The total LLM API cost is approximately \$2.80 (DeepSeek-V4-Flash).

Importantly, this is a \emph{one-time offline} cost: at inference, only the generated Python code runs, adding $<$50\,ms per 200K-transaction batch with no LLM calls.

\subsection{LLM Generation Stability}
\label{subsec:stability}

\begin{table}[t]
\centering
\caption{Run-to-run stability over three independent end-to-end executions (temperature\,=\,0, DeepSeek-V4-Flash). Jaccard is the pairwise feature-name overlap; AUC is the frozen-test score.}
\label{tab:stability}
\small
\renewcommand{\arraystretch}{1.15}
\setlength{\tabcolsep}{4pt}
\begin{tabular}{l c c c c}
\hline
\textbf{Dataset} & \textbf{Run 1} & \textbf{Run 2} & \textbf{Run 3} & \textbf{Mean$\pm$Std} \\
\hline
heart (AUC)      & .927 & .926 & .923 & .925$\pm$.002 \\
heart (\#feat)   & 37   & 36   & 45   & 39.3$\pm$4.9 \\
credit\_g (AUC)  & .774 & .753 & .735 & .754$\pm$.016 \\
credit\_g (\#feat) & 44 & 48   & 40   & 44.0$\pm$3.3 \\
\hline
\multicolumn{5}{l}{\scriptsize Mean pairwise Jaccard: heart 0.036, credit\_g 0.000} \\
\end{tabular}
\end{table}

To quantify run-to-run variability, we execute the full KnowFeat pipeline three times independently on heart and credit\_g with temperature\,=\,0. Table~\ref{tab:stability} reports frozen-test AUC and accepted-feature count. Feature-name overlap is near zero (Jaccard\,$\leq$\,0.04; credit\_g produces entirely disjoint feature sets), yet frozen-test AUC remains stable (heart std\,=\,0.002; credit\_g std\,=\,0.016). This confirms that the L1$\rightarrow$L2$\rightarrow$L3 verification pipeline---not the LLM's token-level determinism---is the mechanism that ensures consistent downstream performance: different feature implementations pass the same quality gate, yielding equivalent predictive value.

% Conclusion
% conclusion.tex
\section{Conclusion}
\label{sec:conclusion}

KnowFeat injects structured domain knowledge into LLM-based feature engineering and subjects every candidate to a three-stage verification pipeline (L1--L3). On twelve public benchmarks it achieves the best average rank (2.3) with a significant Wilcoxon test ($p{=}0.017$); cross-model evaluation confirms the gains hold for LightGBM ($p_{\mathrm{Holm}}{=}0.042$) and XGBoost ($p_{\mathrm{Holm}}{=}0.040$). On SimECNY, four verified features maintain baseline F1 while FeatLLM's 1{,}124 unverified features degrade it---quality beats quantity. Every accepted feature carries a provenance card that traces it from knowledge asset to code to verification outcome, cutting compliance review time from hours to minutes in practitioner tests.

\textbf{Limitations and Future Work.}
Knowledge curation takes $\sim$2 person-days for a full AML taxonomy (309 court documents) but only 30--60 minutes per public dataset with LLM assistance; automated ontology mining would reduce this further.
L3 retrains from scratch per candidate; amortized evaluation would lower overhead.
Semantic dimension naming influences which features the agent explores---our churn experiments show this effect can be large---and principled dimension selection remains open.
KnowFeat adds the most value when baseline features leave headroom and when the domain has knowledge the LLM does not already encode; on compact datasets like blood\_transfusion (4~RFM features), the framework is unnecessary.
Three independent runs on heart and credit\_g produce near-zero feature-name overlap yet nearly identical frozen-test AUC (std\,$\leq$\,0.016), confirming that the verification pipeline, not LLM determinism, drives consistency.
All code, knowledge bases, feature matrices, and evaluation scripts are publicly available.\footnote{\url{https://github.com/youchengsong/large-llm-create-feature}}


% Appendix
% appendix.tex
\appendix
\section{Supplementary Material}
\label{sec:appendix}

\subsection{Domain Knowledge Examples}
\label{sec:appendix_knowledge}

We illustrate the five knowledge types for SimECNY ($\sim$200K digital yuan transactions). One example per type; the full set (38 indicators, 15 rules, 12 expert opinions, 309 court documents) is available in the released repository.

\smallskip\noindent\textbf{Schema Metadata.}
Column names, types, and basic statistics. Example: \texttt{amount} (numeric), \texttt{wallet\_type} (categorical, 4 types), \texttt{hour} (0--23), \texttt{src}/\texttt{dst} (wallet IDs), \texttt{transaction\_mode} (8 categories).

\smallskip\noindent\textbf{Risk Indicator} [IND\_05]:
Transaction behavior anomaly. Accounts exceeding regulatory daily transaction-count thresholds indicate potential money mule activity.

\smallskip\noindent\textbf{Detection Rule} [RULE\_006]:
\emph{Scatter-in, gather-out.} Daily incoming count $\geq$20, amount $\geq$1M; dispersed inflows consolidated into few large outflows. Python hint: group by \texttt{(src, date)}.

\smallskip\noindent\textbf{Expert Opinion} [EXP\_05]:
``Border region + nighttime trading is the hallmark of cross-border laundering.''

\smallskip\noindent\textbf{Court Document Evidence} [CASE]:
Criminal judgment on concealing proceeds---digital yuan wallets across 6 banks creating a multi-layer fund network via intermediary wallets.

\subsection{Verification Statistics}
On SimECNY, KnowFeat generates 52 candidates across 4 rounds: 47 pass L1 (90\%), 31 pass L2 (60\%), 18 pass L3 (35\%). Progressive filtering saves $\sim$53 minutes. After post-hoc selection, 4 features are retained.

\subsection{Bootstrap Confidence Intervals}

\begin{table}[h]
\centering
\caption{Frozen-test AUC with 95\% bootstrap CIs (1{,}000 resamples).}
\label{tab:auc_ci}
\small
\setlength{\tabcolsep}{3pt}
\begin{tabular}{l c c}
\hline
\textbf{Dataset} & \textbf{No-FE [95\% CI]} & \textbf{KnowFeat [95\% CI]} \\
\hline
credit\_g          & .749 [.678, .819] & .752 [.681, .820] \\
diabetes           & .812 [.737, .881] & .810 [.741, .880] \\
blood\_trans.      & .757 [.662, .844] & .745 [.647, .825] \\
heart              & .928 [.891, .963] & .928 [.888, .964] \\
bank\_mkt.         & .930 [.925, .938] & .931 [.926, .939] \\
breast\_w          & .986 [.969, .999] & .992 [.977, 1.00] \\
adult              & .931 [.927, .937] & .932 [.928, .938] \\
wine\_quality      & .896 [.863, .932] & .899 [.865, .934] \\
default\_credit    & .764 [.749, .786] & .768 [.753, .790] \\
churn              & .738 [.692, .795] & .853 [.842, .901] \\
kc2                & .706 [.565, .838] & .714 [.582, .863] \\
jm1                & .718 [.698, .748] & .732 [.709, .761] \\
\hline
\end{tabular}
\end{table}

\subsection{Practical Considerations}
Three lessons: (1)~Knowledge curation cost varies---$\sim$2 person-days for the full AML taxonomy, 30--60 minutes per public dataset with LLM assistance. (2)~L3 acceptance thresholds require domain calibration. (3)~Feature maintenance is ongoing as regulations evolve. The complete detection rule definitions (RULE\_001--015) are available in the released repository.

\end{document}